\documentclass[twocolumn,prx,showpacs,amsmath,amstex,amssymb,mathfonts,superscriptaddress]{revtex4-1}

\pdfoutput=1
\usepackage{amsthm,color,amsfonts,graphicx,verbatim}
\usepackage{amsmath}
\usepackage{amssymb}
\usepackage{amsthm}

\usepackage[normalem]{ulem}

\usepackage{amsfonts}
\usepackage{listings}
\usepackage{enumerate}
\usepackage{latexsym}
\usepackage{bm}
\usepackage{graphicx}
\usepackage[breaklinks,colorlinks = true,linkcolor = red,urlcolor=cyan,citecolor=red]{hyperref}
\newcommand{\beq}{\begin{equation}}
\newcommand{\eeq}{\end{equation}}
\newcommand{\be}{\begin{eqnarray}}
\newcommand{\ee}{\end{eqnarray}}

\newcommand{\bra}[1]{\left\langle{#1}\right|}
\newcommand{\ket}[1]{\left|{#1}\right\rangle}

\renewcommand{\paragraph}[1]{\vspace{4pt}\noindent {\it #1.---}}

\begin{document}
\title{Spin-catalyzed hopping conductivity in disordered strongly interacting quantum wires}
\author{S. A. Parameswaran}
\affiliation{Department of Physics and Astronomy, University of California, Irvine, CA 92697, USA}
\affiliation{Kavli Institute for Theoretical Physics, University of California, Santa Barbara, CA 93106}

\author{S. Gopalakrishnan}
\affiliation{Kavli Institute for Theoretical Physics, University of California, Santa Barbara, CA 93106}
\affiliation{Department of Physics and Burke Institute, California Institute of Technology, Pasadena, CA 91125, USA}

\date{\today}
\begin{abstract}

{In one-dimensional electronic systems with strong repulsive interactions, charge excitations propagate much faster than spin excitations. Such systems therefore have an intermediate temperature range [termed the ``spin-incoherent Luttinger liquid'' (SILL) regime] where charge excitations are ``cold'' (i.e., have low entropy) whereas spin excitations are ``hot.'' We explore the effects of charge-sector disorder in the SILL regime {in the absence of external sources of equilibration}. We argue that the disorder localizes all charge-sector excitations; however, spin excitations are protected against full localization, and act as a heat bath facilitating charge and energy transport on asymptotically long timescales. The charge, spin, and energy conductivities are widely separated from one another. The dominant carriers of energy are neither charge nor spin excitations, but neutral ``phonon'' modes, which undergo an unconventional form of hopping transport that we discuss. We comment on the applicability of these ideas to experiments and numerical simulations.}
\end{abstract}
\maketitle


\section{Introduction}

Interacting electrons in one dimension  behave as \textit{Luttinger liquids}~\cite{HaldaneLuttinger} at low temperature: 
their elementary excitations are  
collective charge- and spin-density waves, which propagate at different velocities---a phenomenon 
 known as ``spin-charge separation''~\cite{giamarchi_book}. When the electron-electron interactions are strong, the spin and charge velocities are widely separated, and for repulsive interactions, charge propagates much faster than spin. Thus, systems with strong repulsive interactions host  an intermediate temperature regime where the temperature is large compared with the kinetic energy of spin excitations but small compared with the kinetic energy of charge excitations. As a consequence, 
 the charge excitations are close to their ground state whereas the spin excitations are essentially at infinite temperature~\cite{berkovich1991, cheianov, matveev2004, FieteBalents}; hence, systems in this regime have been dubbed ``spin-incoherent Luttinger liquids'' (SILLs)~\cite{fiete_RMP}. The equilibrium properties of the SILL are qualitatively different from those of the conventional Luttinger-liquid regime, where both charge and spin sectors are at low temperature~\cite{giamarchi_book}. 
 
Because spin excitations in the SILL regime are effectively at infinite temperature, their equilibrium density matrix is close to the identity, so the \emph{thermodynamic} properties of the SILL are independent of the spin energy scale. 
Although degrees of freedom at infinite temperature do not contribute to \emph{thermodynamics},  
they can still govern \emph{dynamics}. This situation obtains, for example, in disordered \emph{isolated} quantum systems, which undergo a many-body localization (MBL) transition~\cite{BAA, Gornyi} even at infinite temperature~\cite{PhysRevB.75.155111} (see Refs.~\cite{husereview, altmanreview} for recent reviews). We argue here that a similar situation arises in the disordered, isolated SILL: even weak disorder localizes charge excitations, causing the intrinsic charge relaxation timescale to diverge. Instead, the dominant mechanism for charge dynamics involves transitions that borrow energy from the ``hot'' spin bath 
{ --- the spins `catalyze' conduction by placing processes on shell that in their absence would be forbidden due to energy conservation.}
 The SILL regime is an unusual setting for exploring such phenomena: prior studies of MBL have involved systems {\it all} of whose degrees of freedom are cold~\cite{BAA} or hot~\cite{PhysRevB.75.155111}, whereas in the SILL some degrees of freedom are cold and others are hot. 

Understanding transport and relaxation in this regime is important, first, because experimental proposals for realizing the SILL regime (and related systems such as the strong-coupling limit of the Hubbard and $t-J$ models) tend to involve systems that are well-isolated from their environments and temperatures where phonon-mediated relaxation is unimportant. Moreover, relaxation in ``two-component'' systems---involving high-frequency, tightly localized modes coupled to low-frequency delocalized modes---naturally arises in multiple experimental settings. For instance, the experiments of Ref.~\cite{bordia} involve quasi-1D geometries, in which localized longitudinal modes can relax by coupling to delocalized transverse modes {whose bandwidth is tunable by varying lattice depth}; in solid-state systems, nuclear spins can play a similar role in thermalizing the dynamics of electron spins. While such ``narrow-bath'' systems ultimately establish ergodic dynamics, the crossover to such behavior occurs over asymptotically long time scales: the dynamics at shorter times may retain imprints of the (avoided) localized phase, {\it e.g.}, via the parameter-dependence of relaxation timescales~\cite{MeanFieldMBLTransition}. As true many-body localization is an experimentally elusive ideal --- particularly in the solid-state setting --- understanding such dynamical crossovers is an important route to study various intriguing phenomena that have been proposed to occur in the MBL regime and proximate to the localization transition~\cite{2014arXiv1407.7535B,Agarwal,VHA, PVPtransition, luitz2016}.

Accordingly, here we advocate  that the disordered SILL is profitably viewed as an instance of a ``nearly many-body localized metal''~\cite{MeanFieldMBLTransition}, and consequently studying its transport properties may provide insight into universal properties of many-body localized systems in one dimension. 
 Specifically, we argue that transport in isolated disordered wires in the SILL regime is governed by the small spin energy scale, since in the absence of the thermalizing spin bath, the system is (many-body) localized. 
 We consider a hierarchy of scales in which the charge energy scale is the largest in the problem, followed successively 
 by the disorder strength, the temperature, and the spin energy scale. In this regime, disorder localizes the low-energy charge excitations~\footnote{In principle, high-energy charge excitations that lie outside our effective theory can provide activated transport; at the temperatures of interest, however, this mechanism is subleading to what we consider.}. Thus, charge excitations on their own do not give rise to transport in the d.c. limit. However, the spin excitations act as a slowly fluctuating thermal bath (which is protected from localization by the $SU(2)$ spin-rotation symmetry, provided that spin-orbit coupling is absent). Charge and energy transport then take place through various forms of variable-range hopping mediated by this slowly fluctuating internal spin bath.  { The resulting energy transport is parametrically faster than charge transport, but both rates show a non-trivial power-law dependence on the spin energy scale. We discuss how this spin-catalyzed hopping conductivity  can be tested in both cold-atom and solid-state experiments as well as in numerical simulations.}

Before proceeding, we place the present paper in the context of other related work. Previous investigations of transport in SILLs~\cite{fiete_transport} have focused on single-impurity problems, rather than the case of a finite density of quenched impurities that is pertinent to localization physics. Hopping conductivity (both d.c. and a.c.) in Luttinger liquids has been recast in terms of effective two-level systems~\cite{malininVRH,rosenow_nonlinear} and pinned charge-density waves~\cite{GiamarchiSchulz, larkinlee, feigelman80, feigelman81, ledoussal}, but those prior works all assumed the existence of a ``perfect bath'' capable of placing any hopping process on-shell; this is in marked contrast to the narrow-bandwidth bath, natural in the SILL context, that we consider here. The phenomenology of ``narrow bath'' disordered systems was studied in ~\cite{MeanFieldMBLTransition}  but there the focus was on developing a mean-field approach to the MBL transition, rather than on transport properties. Furthermore, in contrast to many analytical treatments of MBL that work in the limit of a weakly interacting Anderson  (i.e., free-fermion) insulator, the approach here builds in strong interactions at the outset. Finally, we note that while variable-range hopping conductivity has an extensive history, our computation of the thermal conductivity $\kappa$ of the SILL (Sec.~\ref{sec:energy}) is quite unusual, and to our knowledge has no antecedents in the literature. {The dominant thermal conduction channel is the hopping of {\it bosonic} excitations above the SILL ground state, corresponding to  Gaussian fluctuations of the localized charge degrees of freedom viewed as a classically pinned charge-density wave. (Charge transport, in contrast, occurs due to instanton events tunneling between near-degenerate classical configurations, as the Gaussian sector is neutral.) Given a perfect bath, the contribution of these modes to $\kappa$ would be negligible as a local distortion of these modes (whose number is not conserved) would decay before it can hop,} 
but 
energy conservation combined with the narrow bath bandwidth forbids this. As a consequence, the Gaussian sector exhibits an emergent conservation law: the total occupation of modes within a bath bandwidth is approximately constant, permitting a `foliation' of the spectrum into a set of narrow thermal conducting channels; optimizing over these channels then produces a contribution to $\kappa$ that dominates that  of both the instantons and of the spins in the bath themselves.

The rest of this paper is structured as follows. In Sec.~\ref{sec:SILL} we specify our model and review standard results on the SILL regime. In Sec.~\ref{sec:scales} we provide a heuristic discussion of the dynamics of the disordered SILL regime. In Sec.~\ref{sec:charge} we discuss both a.c. and d.c. charge conductivity. In Sec.~\ref{sec:energy} we estimate the d.c. thermal conductivity, which we argue is parametrically larger than the d.c. charge conductivity (being due to a different physical channel.) Finally, in Sec.~\ref{sec:disc} we summarize our results and discuss the effects of phonons and spin-orbit coupling, as well as implications for experiment.


{
\section{Bosonized Effective Theory and Microscopic Models} \label{sec:SILL}
}
\subsection{Universal description in the spin-incoherent regime}

We begin by introducing the general effective Hamiltonian that describes the spin-incoherent Luttinger liquid (SILL) regime; specific microscopic realizations are discussed in Sec.~\ref{parents}. This Hamiltonian consists of three parts, $H = H_c + H_s + H_{\text{irr}}$. The charge excitations are described by~\cite{giamarchi_book}
\be
H_c = \frac{v_c}{2\pi} \int dx \left\{ K_c [\partial_x \theta_c(x)]^2 + \frac{1}{K_c} [\partial_x \phi_c(x)]^2 \right\}.
\ee
Here, the bosonic fields $\theta_c, \phi_c$ obey the canonical commutation relations $[\theta_c(x), \phi_c(y)] = -i\frac{\pi}{2}\text{sgn}(x-y)$ \footnote{Throughout, we use the standard conventions of \cite{GiamarchiSchulz} and work in $\hbar=k_B=e=1$ units.}. The spin Hamiltonian $H_s$ is taken to have some generic local lattice form, in terms of local operators $h_i$:

\be
H_s \simeq W_s \sum_i h_i,
\ee 
 where the $h_i$ are chosen so that $H_s$ is invariant under $SU(2)$ rotations. In the SILL regime, the overall energy scale $W_s$ is small enough that $\exp(-H_s / T) \approx 1$.  Thus, the spin Hamiltonian does not affect thermodynamics or equilibrium properties in this regime---a feature known as ``super-universality''~\cite{fiete_RMP}. Since we are interested in \emph{dynamics} as well as thermodynamics, we specify that the long-time autocorrelation functions of the spin Hamiltonian follow linearized hydrodynamics. Thus, for example, 
\be
\langle S^x_i(t) S^x_i(0) \rangle \sim {1/\sqrt{\mathcal{D}{t}}}
\ee
where $\mathcal{D} \sim W_s$ is the spin diffusion constant. Additionally, non-conserved operators will decay exponentially with a rate that is similarly set by $W_s$. These assumptions would hold, in particular, if the high-temperature dynamics of $H_s$ were thermal {and ergodic}, as they generically will be.

The spin-charge coupling is given by a generic $SU(2)$-symmetric form, such as
\be
H_{\text{irr}} \simeq g \sum_i \int dx h_i [\partial_x \phi_c(x)]^2 \delta(x - x_i),
\ee
where $x_i$ is the position of the $i$th lattice site. The spin-charge coupling is irrelevant in the renormalization-group sense; however, for our purposes it is \emph{dangerously} irrelevant, as without it the spin and charge sectors would not equilibrate with each other. 
Recall that we are interested in the \emph{unitary} dynamics of an \emph{isolated} quantum system governed by the Hamiltonian $H$; thus, the couplings in $H_{\text{irr}}$ will be crucial ingredients in the relaxation times we compute.

A key characteristic of the SILL is {\it superuniversal} spin physics: regardless of the strength of the spin-spin interactions, at energy scales large compared to $W_s$, the spin dynamics drop out of the problem and the theory is effectively spinless. To see why, consider working in the regime $W_s\ll T\ll W_c$. Then, over the typical thermal coherence time $\tau_{\text{th}} \sim 1/T$, the spins are effectively frozen: as their typical time scale is $\tau_s \sim 1/W_s$, the probability of a transition between distinct spin states is negligible on a time scale $\tau_{\text{th}}$. Note that on this same time scale, {\it charges}  can fluctuate dynamically, since $\tau_{c}\sim W_c^{-1} \ll \tau_{\text{th}}$. Crucially, this remains within the regime of applicability of the low-energy theory $H$, allowing us to use Luttinger liquid techniques. The static nature of spins over timescales $t \ll\tau_{\text{th}}$ allows us to effectively strip the spins from the charges, and replace the SILL by an effective spinless LL with twice the density, and hence a doubled Luttinger parameter  $K_\text{eff}= 2K_c$~\cite{fiete_transport}, where $K_c$ is the Luttinger parameter for charge.

The properties of the SILL at energy scales much greater than $W_s$ can therefore be computed by mapping it onto the effective Hamiltonian
\be\label{heff}
{H}_{\mathrm{eff}} = \frac{v_c}{2\pi} \int dx \left\{ 2K_c [\partial_x \theta(x)]^2 + \frac{1}{2K_c} [\partial_x \phi(x)]^2 \right\},
\ee
where we have introduced $\theta$, $\phi$, the bosonic fields of the effective spinless theory. This effective model will be central to the remainder; in the rest of this section, we sketch possible microscopic origins for $H_{\text{eff}}$.

\subsection{Microscopic Parent Hamiltonians}\label{parents}

\subsubsection{Hubbard and $t-J$ models}

While we anticipate that our results apply to any system that is described at low energies by $H$ with $v_s\ll v_c$, for concreteness we note that one specific example is the large repulsive $U$ limit of the fermionic Hubbard model,
\be\label{eq:HubbardH}
H_{\text{Hub}} = -t\sum_{i,\sigma=\uparrow\downarrow} \left(c^\dagger_{i\sigma}c^{\phantom\dagger}_{i+1\sigma} +\text{h.c.}\right) + U\sum_i n_{i\uparrow}n_{i\downarrow},
\ee
where $c^\dagger_{i\sigma}$ creates a fermion with spin $\sigma$ on site $i$. In the limit $t\ll U$,  we may impose the constraint of no double occupancy (except in virtual processes), allowing us to reduce (\ref{eq:HubbardH}) to the $t-J$ model,
\begin{align}\label{eq:tJH}
H_{t-J} =&-t\sum_{i,\sigma=\uparrow\downarrow} \left(c^\dagger_{i\sigma}c^{\phantom\dagger}_{i+1\sigma} +\text{h.c.}\right)\nonumber\\&+ J\sum_i \left(S_{i}S_{i+1} -\frac{1}{4} n_i n_{i+1}\right) +\ldots,
\end{align}
where the $S_i$ are spin operators, and $J= 4 t^2/U\ll t$. The ellipsis `$\ldots$' represent higher-order terms conventionally ignored in the $t-J$ limit, but necessary to break integrability within the spin sector (such as second-neighbor terms at order $t^3/U^2$) in order to thermalize the spins in isolation. It can be shown that the  low-energy description of (\ref{eq:tJH}) takes the form of $H$ with $v_s/v_c\sim t/U \sim J/t \ll 1$. {We note that the $t-J$ model has been used in time-dependent density-matrix renormalization group computations of the electronic spectral function in the SILL regime~\cite{feiguin-fiete}, and is perhaps a promising starting point for numerical simulations of the disordered limit of the SILL studied in this paper.}

\subsubsection{Fluctuating Wigner solid}

A different microscopic starting point~\cite{fiete_RMP}, which is more natural for low-density electron gases, is the fluctuating Wigner solid (or harmonic-chain) model. This has the microscopic Hamiltonian

\be\label{hws}
H_{\text{WS}} = \sum_{i = 1}^N \frac{p_i^2}{2M} + \frac{M \omega_0^2}{2} (x_i - x_{i + 1})^2
\ee
where $M$ is a particle mass, and $(x_i, p_i)$ are position and momentum coordinates of the $i$th particle. In two or more dimensions, this model has a crystalline phase at zero temperature; in one dimension, however, crystalline order is forbidden even at zero temperature. Nevertheless, in the limit of large $M$, the typical root-mean-squared displacement of a particle (in the background created by the other particles) is much less than the interparticle spacing. Thus, exchange effects (which set the spin energy scale) are suppressed. Consequently, the spin energy scale is once again parametrically smaller than the charge energy scale, which is set by $\omega_0$. 

A feature of the Wigner solid model that will be helpful for our purposes is that it has a well-defined classical limit ($M \rightarrow \infty$), in which kinetic terms are absent and the system is a classical charge-density wave. This limit corresponds to the Luttinger parameter $K_\rho \rightarrow 0$. By contrast, in the Hubbard and $t-J$ models, this semiclassical limit is absent: even when $U \rightarrow \infty$, the kinetic energy is not quenched, and the dynamics do not become classical. 


{
\section{Adding Disorder to the SILL}\label{sec:scales}}
Having provided an overview of the features of SILL physics in the clean limit, we now perturb our effective theory by introducing charge-sector disorder~\cite{fiete_transport}, which couples to the effective bosonic fields via
\be
H_{\mathrm{dis}} \sim D \int dx\, \xi(x) e^{i \phi(x) + i k_F x} + \mathrm{h.c.}
\ee
Here, $D$ is the characteristic disorder energy scale, while $\xi$ is a random variable such that $\langle \xi \rangle = 0$ and $\langle \xi(x) \xi(x') \rangle \sim \delta(x - x')$. For our purposes we require that $D > W_s$, the spin bandwidth; then, for reasons noted in the preceding section, it is appropriate to use ${H}_{\mathrm{eff}}$ (with the doubled Luttinger parameter $K_{\text{eff}}$) to study disorder effects. 
\\
\subsection{Pinning, Localization, and Length Scales}

A Luttinger liquid description such as $H_{\text{eff}}$ foregoes a single-electron description in favor of  one in terms of the dominant collective degrees of freedom --- in the spinless case, this corresponds to an incipient charge density wave (CDW) oscillating at $2k_F$ ({where $k_F = 2\pi/(n_\uparrow + n_\downarrow)$ is the Fermi wavevector of the spinless model}). As true long-range order is precluded in one dimension by the Hohenberg-Mermin-Wagner theorem, absent disorder, the CDW is (at best) quasi-long-range ordered, with algebraic decay of density-density correlations. Introducing disorder {pins} the CDW, leading to exponential decay of these correlations at a {\it pinning length} $\xi_p$, that may be estimated as follows. 
As the disorder couples directly only to charge, we can invoke the standard analysis of disorder in Luttinger liquids~\cite{GiamarchiSchulz}, to conclude that disorder is relevant, and pinning of the CDW sets in, whenever $K_{\text{eff}}<3/2$. Since we are considering strongly repulsive fermions, we may assume $K_c<3/4$ so that $K_{\text{eff}} < 3/2$ and hence disorder is always relevant in the regime of interest. The arguments of Ref.~\cite{GiamarchiSchulz, feigelman80, feigelman81} then suggest that the pinning scale is given by $\xi_p \sim 1/D^{1/(3 - K_{\text{eff}})}$. However, note that pinning is essentially a {\it classical}  effect --- underscored by the finiteness of $\xi_p$ even in the $K_{\text{eff}}\rightarrow0$, classical, limit. Quantum (or indeed, thermal) fluctuations serve primarily to renormalize parameters, but are not {\it fundamentally} necessary for pinning to occur. More ``quantum'' aspects of localization are reflected in the tunneling between degenerate classical configurations that is, for example, responsible for hopping conductivity of charge in the presence of a bath.

{In describing possible excitations of a pinned CDW, it is instructive to begin in this classical limit}~\footnote{Note that this limit is not always accessible in a given microscopic realization, {\it e.g.} in the Hubbard or $t-J$ models $K_{\text{eff}} \lesssim 1$ always; for concreteness, it may help to consider the  fluctuating Wigner-solid model discussed in Sec.~\ref{parents}, where  the classical limit  corresponds to quenching the kinetic energy by taking $M \rightarrow \infty$.}. For $K_{\text{eff}}\rightarrow 0$ charges are arranged in their lowest-energy classical configuration~\cite{feigelman80}. Low-energy quantum fluctuations about this configuration for small but nonzero $K$ are of two kinds: (i)~oscillations of a particle about its classical position (the ``Gaussian'' {or ``phonon''} sector), and (ii)~tunneling events between nearly degenerate classical configurations (the ``instanton'' sector~\cite{larkinlee}). 

{Low-energy excitations in both sectors are localized, but as we now argue they have different characteristic localization lengths~\cite{mpv}. It is helpful to think of the system as consisting of randomly coupled segments of clean CDW, each of size $\sim \xi_p$. Gaussian-sector excitations (which involve oscillations with characteristic single-particle displacements {$l_\text{osc}$} much smaller than the interparticle spacing) are correlated over distances $\sim \xi_p$; this is their characteristic localization length. {[As these are phonons of the CDW and disorder explicitly breaks translational symmetry, they are not protected against localization at any energy --- and hence the conclusions of ~\cite{AltmanMarginalPhononVRH} do not apply here.]} Instanton-sector excitations, by contrast, involve charge motion over distances that are large compared with a lattice spacing (and, in the regimes we shall focus on, large compared with $\xi_p$ as well). Consider an instanton that moves charge between two nearly degenerate positions separated by a distance $L$. This process involves tunneling through a barrier with a width $\sim L$ and a height that depends on the interaction strength, and its matrix element is thus suppressed exponentially in $L$, with a coefficient that vanishes in the classical limit. We define a ``quantum length'' $\xi_q$ through the condition that the instanton matrix element falls off as $\exp(-L/\xi_q)$. In general, $\xi_q \ll \xi_p$ whenever $K_{\text{eff}} \ll 1$~\footnote{The existence of multiple relevant length-scales is pointed out in Ref.~\cite{mpv}, and relations between the various length-scales are derived there. The length-scale we term $\xi_q$ is related to the scale labeled $\xi_{\mathrm{tun.}}$ in that work. For our purposes, it will suffice to treat these two relevant length-scales as separate parameters.}}

 \begin{figure}[t]
\begin{center}
\includegraphics[width = 0.45 \textwidth]{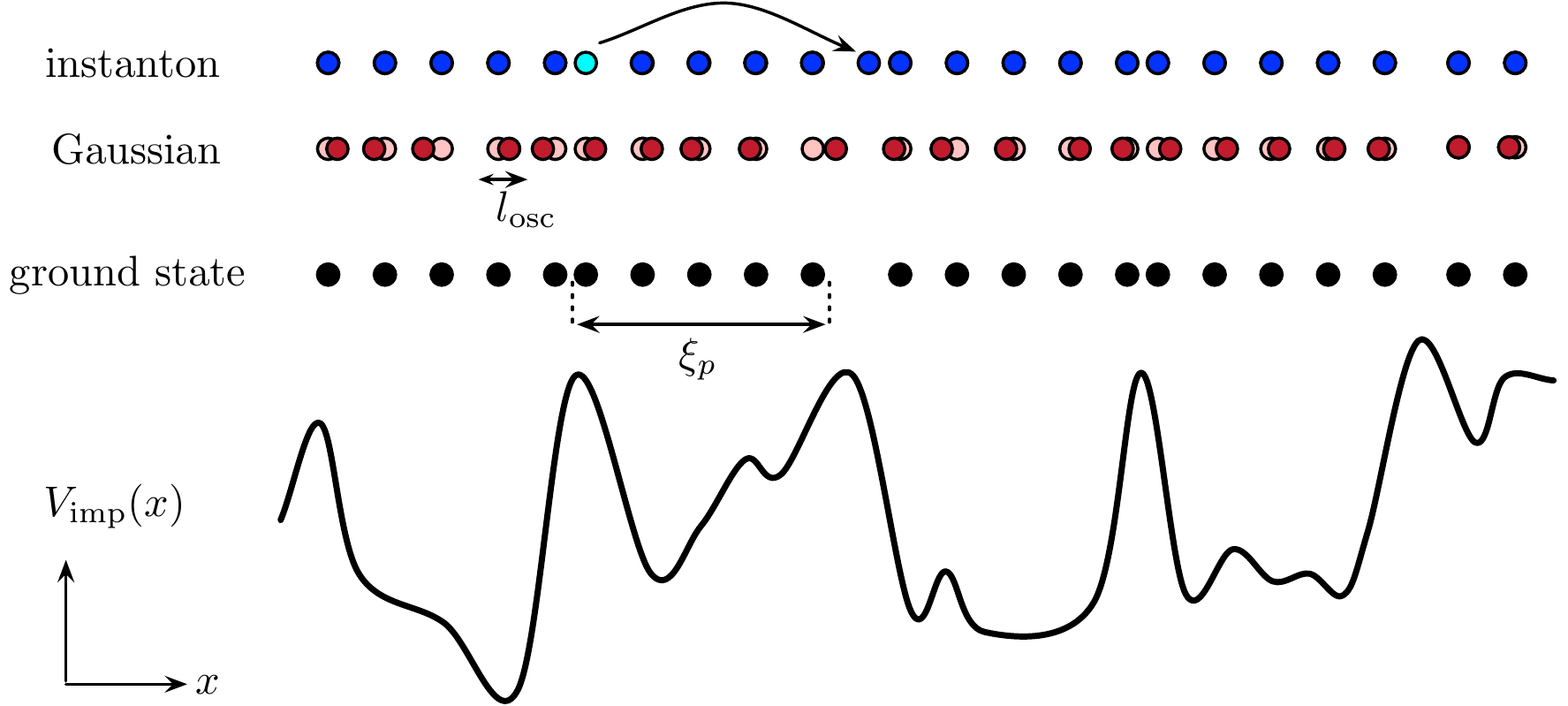}
\caption{{Gaussian modes versus instantons.  
In the presence of an impurity potential $V_{\text{imp}}(x)$, near the classical ($K\rightarrow 0$) limit the ground state configuration of the SILL is a pinned CDW (black dots), retaining short-range density-wave order on scales of order the pinning length $\xi_p$. Excitations of this pinned CDW,  may be divided into (i)~Gaussian (quantum or thermal) fluctuations of the charges about their equilibrium configuration (red dots) with amplitudes $l_\text{osc}$ much smaller than the inter-particle spacing, and (ii)~instanton events  (blue dots) that describe quantum tunneling between nearly degenerate classical saddle points. The instantons involve large-scale charge rearrangements, whereas the Gaussian modes can transmit energy, but not charge, over long distances. All hopping processes of relevance to transport occur on asymptotically longer scales, as discussed in the main text.}
}
\label{fig:pinning}
\end{center}
\end{figure}

This separation of scales between $\xi_p$ and $\xi_q$ implies that there are two typically small parameters in the SILL regime: namely, the Luttinger parameter $K_{\text{eff}}= 2 K_c$ {(which quantifies the ``semiclassicality'' of the dynamics)}, and the ratio of spin to charge bandwidths, $W_s/W_c$. The relation between these is not universal. For example, in the Hubbard model~\cite{schulz_hubbard_1990} at $U/t \gg 1$, the ratio of spin to charge bandwidths vanishes as $t/U$, but $K_c \simeq 1/2$. For a one-dimensional Wigner crystal, $K_c$ vanishes algebraically with the dimensionless interparticle spacing $r_s$~\cite{meyer2008wigner}, whereas $W_s/W_c$ vanishes~\cite{pivovarov} as $\exp(-\eta \sqrt{r_s})$, for some constant $\eta$.

\subsection{Properties at finite energy density}

In the previous subsection we introduced the two kinds of low-energy excitations of the pinned CDW: approximately Gaussian phonons with a localization length $\xi_p$ and instantons (which are non-local two level systems that one can regard as fermions) that have a localization length $\xi_q$. We now consider how the properties of these different excitations are affected at low but finite energy density.

\textbf{\textit{Gaussian sector}.} The Gaussian sector consists of bosonic modes at frequency $\omega_p$, generically with anharmonicities; at finite temperature these modes will be thermally occupied. One can partition these bosonic modes into ``classical'' modes (for which $T \gg \omega_p$,  
with occupation $\sim T/\omega_p$~\footnote{This is likely an overestimate because anharmonicities will tend to limit mode occupancy. We will see later that, even overestimating their occupation, these modes do not dominate response.}), and ``quantum'' modes, which are close to their ground state, and have occupancy $\sim \exp(-\omega_p/T)$. The density of states goes as $\omega_p^3$ at low frequencies~\cite{fogler}; we assume that the temperature is such that all relevant modes are in this low-frequency tail. 

\textbf{\textit{Instanton sector}.} Instantons with a splitting much smaller than $T$ are essentially at infinite temperature, whereas those with splitting much larger than $T$ are in their ground state. Our interest is mainly in the low-frequency limit $\omega_i \ll T$, so the instantons with splitting $\omega_i \agt T$ will be mostly irrelevant to our analysis.

When interaction effects are absent, 
therefore, the relevant degrees of freedom are a thermally occupied ensemble of localized bosonic modes (with localization length $\xi_p$) and localized fermionic modes (with localization length $\xi_q \ll \xi_p$). Adding interactions at finite temperature alters this picture in {three} ways. {First, interactions couple the localized low-energy excitations of the CDW to high-energy charge modes (with energies $\sim W_c$), which are presumably delocalized in the weakly disordered, strongly interacting limit of interest to us (but see Refs.~\cite{modak_mobedge, pixley_mobedge}). These delocalized modes can transport charge and act as a bath for the low-energy sector. However, their effects are suppressed by the Boltzmann factor $\exp(-W_c/T)$, and will be subleading in the $W_c \ll T$ regime of interest to us.}
{Second}, interactions permit many-body resonances, involving the simultaneous rearrangement of several particles~\cite{mbmott, jzi, abanincriterion, rademaker}; such processes are absent in the ground state (which is unique) but possible in thermal states (which have a finite entropy), and will be relevant to our discussion of transport below. {Third}, it is in principle possible for very low frequency classical modes from the Gaussian sector to delocalize the system, because they may act as local low-frequency drives. Before proceeding, we must ensure that this situation does not arise in the SILL in the dynamical regimes of interest to us.

\textbf{\textit{Stability against Gaussian-sector driving}.} That these spatially sparse classical modes do not delocalize the rest of the system can be seen by the following heuristic argument. The criterion for a drive at strength $A$ and frequency $\omega$ to cause delocalization~\cite{adh2014} is that 

\beq\label{dima}
\left(\frac{A}{\omega}\right)\left(\frac{A}{W_c}\right)^{2/(\xi_p \ln 2)} \agt 1. 
\eeq
Because the classical modes are themselves localized, at a distance $x$ from such a classical mode, the amplitude of its coupling to other modes (and thus the effective drive amplitude) is $A\sim W_c \exp(-x/\xi_p)\sqrt{T/\omega}$ (the factor of $\sqrt{T/\omega}$ is due to Bose enhancement). The spacing between classical modes of frequency $\omega$ is set by $x(\omega) \sim (W_c/\omega)^3$, using the estimate for the tail density-of-states. In order for the rare localized modes at frequency $\omega$ to delocalize the entire system, the criterion~\eqref{dima} would have to be satisfied at distances of order $x(\omega)$, so that each mode can localize the region around it. This would require that
\beq
\frac{W_c}{T} \left(\frac{T}{\omega}\right)^{\frac{3}{2} + \frac{1}{\xi_p \ln 2}} \exp\left[- \left(\frac{W_c}{\omega} \right)^3 \left( \frac{1}{\xi_p} + \frac{2}{\xi_p^2 \ln 2} \right) \right] \agt 1,
\label{eq:classdrivbound}\eeq
which is clearly not the case for sufficiently small $\omega$. Thus, rare classical modes might cause some degree of delocalization in their immediate surroundings, but do not delocalize the entire system. (We emphasize that (\ref{eq:classdrivbound}), which does not consider anharmonicity and treats the modes as purely classical, \emph{overestimates} the extent of delocalization due to these classical modes.)

\subsection{Delocalization via spin bath}

So far, we have ignored the spin degree of freedom completely, and have found that under this assumption the system is effectively localized at finite temperature (up to timescales of order $\exp(-W_c/T)$). Since charge and spin are only weakly coupled, and the disorder does not directly couple to the spin, we expect that the spin sector is thermal, and acts as a weak, slowly fluctuating bath for the charges. We shall assume that the spin-charge coupling $g$ is weak relative to the spin bandwidth $W_s$. In this limit, the spin bath can be treated perturbatively. We emphasize, however, that because of the $SU(2)$ symmetry of the spin sector, spin excitations never freeze out, and some transport is present even in the limit of $g \gg W_s$. {[Note that this conclusion will be altered if $SU(2)$ symmetry is broken, {\it e.g.} by  spin-orbit coupling: in this case, the localized charge distribution can induce a  site-dependent random field on the spin sector, and such back-action may localize the bath --- a so-called ``MBL proximity effect''~\cite{MBLprox}. In the $SU(2)$ symmetric case, only bond disorder is induced on the spins, and this is believed to be robust against MBL~\cite{QCGPRL}.]}

Owing to the scale hierarchy $W_s \ll T$, the manner in which the spin bath delocalizes the charge sector is an unusual form of variable-range hopping. Because the spin sector can only absorb energies smaller than $W_s$, the transitions that it mediates involve pairs of states that are within $W_s$ of one another in energy. The effects of such coupling on relaxation in the instanton sector were previously addressed in Ref.~\cite{MeanFieldMBLTransition}; below, we generalize these results to transport. The effects of $W_s \ll T$ on the \emph{Gaussian} sector, however, are more unusual. Here, the combination of localization and the narrow spin bath gives rise to an approximate boson number conservation: although bosons can be created or destroyed, it takes an energy $\sim T$ to create and destroy them, so the relevant process only takes place at order $T/W_s \gg 1$ in the spin-charge coupling. The dominant channel by which phonons equilibrate, instead, is by hopping between approximately degenerate modes. 
{This is related to a peculiar feature of the spin ``bath'': {namely, that} its heat capacity is far lower than that of the charge sector. {As a consequence}, the relaxation of a nonequilibrium charge configuration does not appreciably change the energy of the spin sector: rather, the spin sector primarily ``catalyzes'' the spreading of energy within the charge sector, by permitting charge transitions that would not otherwise be on shell.}

We have now set the stage for our main discussion: in the next two sections, we will consider charge and energy transport through hopping processes of the Gaussian and/or instanton sectors of the charge modes,  that are placed on-shell by rearrangements of the thermalizing spin bath. 

\section{Charge transport}\label{sec:charge}

In this section we discuss charge transport in the disordered SILL. We begin by discussing the isolated-system result for linear-response charge conductivity due to the instanton sector. We then turn to saturation effects induced by the spin bath, and then finally to conductivity in the d.c. limit. Our discussion of a.c. response---which does not involve the spin bath---is similar in spirit to previous work~\cite{mbmott, rosenow_nonlinear}; however, in the d.c. limit the narrow-band nature of the spin bath leads to striking deviations from the standard hopping-transport predictions~\cite{ahl, sebook}.

\subsection{Optical conductivity in the isolated system}
We begin by considering the optical charge conductivity (ignoring the spin degree of freedom). 
{For this purpose, it is convenient to begin with the Kubo formula,}
\be
{ \sigma(\omega) = \frac{1 - e^{-\omega/T}}{ \omega {Z} N} \sum_{m,n} e^{-E_m/T} \left|\bra{m} j \ket{n}\right|^2 \delta(\omega-\omega_{mn}),} \nonumber
\ee
where ${Z}$ is the partition function, $N$ the number of sites in the system, the indices $m, n$ run over all the many-particle eigenstates, whose splitting is given by $\omega_{mn}$, and the current $j$ is the sum over local currents, $j = \sum_i j_i$. 

{We are interested in the frequency regime $W_s \ll \omega \ll T \ll W_c$. Thus, we can approximate $1 - e^{-\omega/T} \approx \omega/T$.  The Boltzmann factors ${e^{-E_m/T}/Z}$ determine a density per site $\sim T/W_c$ of relevant initial states.
In this regime, and in the absence of interactions, the dominant contribution to the optical conductivity comes from two-level systems (TLSs) consisting of two-site resonances with a splitting that matches the drive at frequency $\omega$. The optical conductivity due to these was derived by Mott~\cite{mott1968}, whose argument we briefly review for completeness. The characteristic size of resonant pairs with splitting $\omega$ is $r_\omega$, determined by the condition $W_c e^{-r_\omega/\xi_q} \sim \omega$; the {current} matrix element of the drive coupling these pairs is $j \sim \omega r_\omega$~\cite{mott1968}. Finally, the phase space of final states goes as $r_\omega^{d - 1}{\xi_q/W_c}$ in $d$ dimensions (and is therefore constant in one dimension). Combining these expressions, we recover the standard expression}

\be\label{spmott}
\sigma_{\mathrm{sp}} (\omega) \sim \left( \frac{\omega}{W_c} \right)^2 {\xi_q^3} \log^2\left(\frac{W_c}{\omega}\right). 
\ee 
{At finite temperature, in the presence of interactions, this expression is modified because multiparticle rearrangements become possible~\cite{mbmott}}. 
{While we may use a similar argument to the single-particle result above, there are important modifications. We must consider resonances that involve many-particle rearrangements, rather than single-particle hops: hence, while there is a formal similarity with the preceding discussion, the variable to be optimized is not the {distance} between single-particle orbitals, but the number of particles rearranged in going between the two eigenstates involved in the transition.  The phase space is thus greatly increased: specifically, the number of possible $n$-particle rearrangements involving a particular particle goes as $e^{sn}$ where $s \simeq T/W_c$ is an entropy density per site (one can also think of $s$ as the density of excited sites). Since the excitation density is low ($\sim s$) the interactions between excitations are also weak in the low-temperature limit: the tunneling matrix element for a two-particle rearrangement will fall off as the wavefunction overlap between the two localized orbitals at a distance $1/s$, which is $\exp(-1/(s \xi_p))$~\footnote{It might not seem immediately apparent why $\xi_p$ rather than $\xi_q$ is the relevant length-scale here. The key observation is that an instanton causes a \emph{distortion} of the CDW with correlation length $\xi_p$, which is what mediates the interaction.}. Thus an $n$-particle rearrangement has tunneling matrix element $W_c\exp(-n/(s \xi_p))$ (replacing the single-particle result $W_ce^{-r/\xi_q}$). Using the Mott criterion, the optimal rearrangements at frequency $\omega$ involve $n_\omega \simeq s \xi_p \log(W_c/\omega)$. The {\it current} matrix elements that enter the Kubo formula retain their dependence on $\omega$ (upto logarithmic factors), so that, upon including the many-body phase space factor for the optimal rearrangements, we find 
}
\be
{\sigma_{\text{int}}(\omega) \sim \left(\frac{\omega}{W_c} \right)^{2 - \gamma \xi_p (T/W_c)},}
\ee
where $\gamma$ is a numerical factor of order unity. Note that these interaction effects are only relevant at sufficiently low frequencies, $\omega \alt W_c e^{-W_c/(T \xi_p)}$. At higher frequencies, the many-body resonances giving rise to Mott-type conductivity are absent, and the single-particle result~\eqref{spmott} applies. 

Coupling to the spin bath does not appreciably change this linear-response result in the regime $W_s \ll \omega \ll T$. However, it does affect the nature of the \emph{steady-state} response~\cite{rosenow_nonlinear}. When dissipation is absent, linear response only occurs as a transient, on timescales short compared with the field amplitude $t \alt 1/(\xi_q E)$. On longer timescales, all the instantons are saturated and there is no further response~\cite{gkd2016, kozarzewski2016, rehn2016}. However, in the presence of a relaxation timescale $\tau$ (which we will estimate below), the steady-state conductivity is given by~\cite{rosenow_nonlinear}
\beq
\sigma_{\mathrm{ss}}(\omega) \simeq \sigma_{\mathrm{int}}{(\omega)} \left[1 - \left( \frac{E \xi_q \log(W_c/\omega)}{1/\tau} \right)^2 \right]
\eeq

\subsection{Relaxation in the presence of the spin bath}

Before turning to the dc conductivity (which is governed by hopping processes that the spin bath mediates), we briefly discuss the relaxation time of a particular two-level system in the presence of the spin bath. This discussion is a straightforward application of the ideas in Ref.~\cite{MeanFieldMBLTransition}. In order for a particular system configuration to relax, it must borrow energy from the spin bath. The spin bath can only contribute $W_s$ of energy to a transition, whereas the typical detuning for nearest-neighbor hops is of order $W_c$. Thus, the size of the region that must be rearranged to find a hop that can be put on shell goes as $l \sim 1/s \log(W_c/W_s)$, {where as before $s\sim T/W_c$ is the density of excited charge sites}. The matrix element for such a transition is $g \exp(-l/\xi_p)$, where $g$ is the spin-charge coupling. Applying the Golden Rule, {using the fact that the density of final states is $\sim 1/W_s$ set by the spin bath, we obtain the rate} 

\beq\label{gammaint}
\Gamma_{\mathrm{int}} \simeq \frac{g^2}{W_s} \left(\frac{W_s}{W_c} \right)^{2 W_c / (\xi_p T)}.
\eeq
Note that the temperature-dependence is \emph{activated}. 

At low temperatures, a competitive process involves single-excitation hopping. Again, this process is bottlenecked by the small spin energy scale: pairs of sites with energy mismatch less than $W_s$ are spaced apart by $r\sim W_c/W_s$, and therefore the matrix element for hopping from one to the other goes as $g \exp[-W_c/(\xi_q W_s)]$. The associated rate is

\beq\label{gammasp}
\Gamma_{\mathrm{sp}} \simeq \frac{g^2}{W_s} e^{-2W_c/(\xi_q W_s)}.
\eeq
Note that this is temperature-independent, so one might expect it to dominate in some temperature regimes. Comparing Eqs.~\eqref{gammaint} and \eqref{gammasp} one finds that interacting processes dominate when

\beq
T \agt W_s \log(W_c/W_s)
\eeq
while single-particle hops dominate relaxation (and the relaxation rate thereby becomes temperature-independent) in the window $W_s \ll T \ll W_s \log(W_c/W_s)$. 

\subsection{Hopping conductivity}

\begin{figure}[t]
\begin{center}
\includegraphics[width = 0.45 \textwidth]{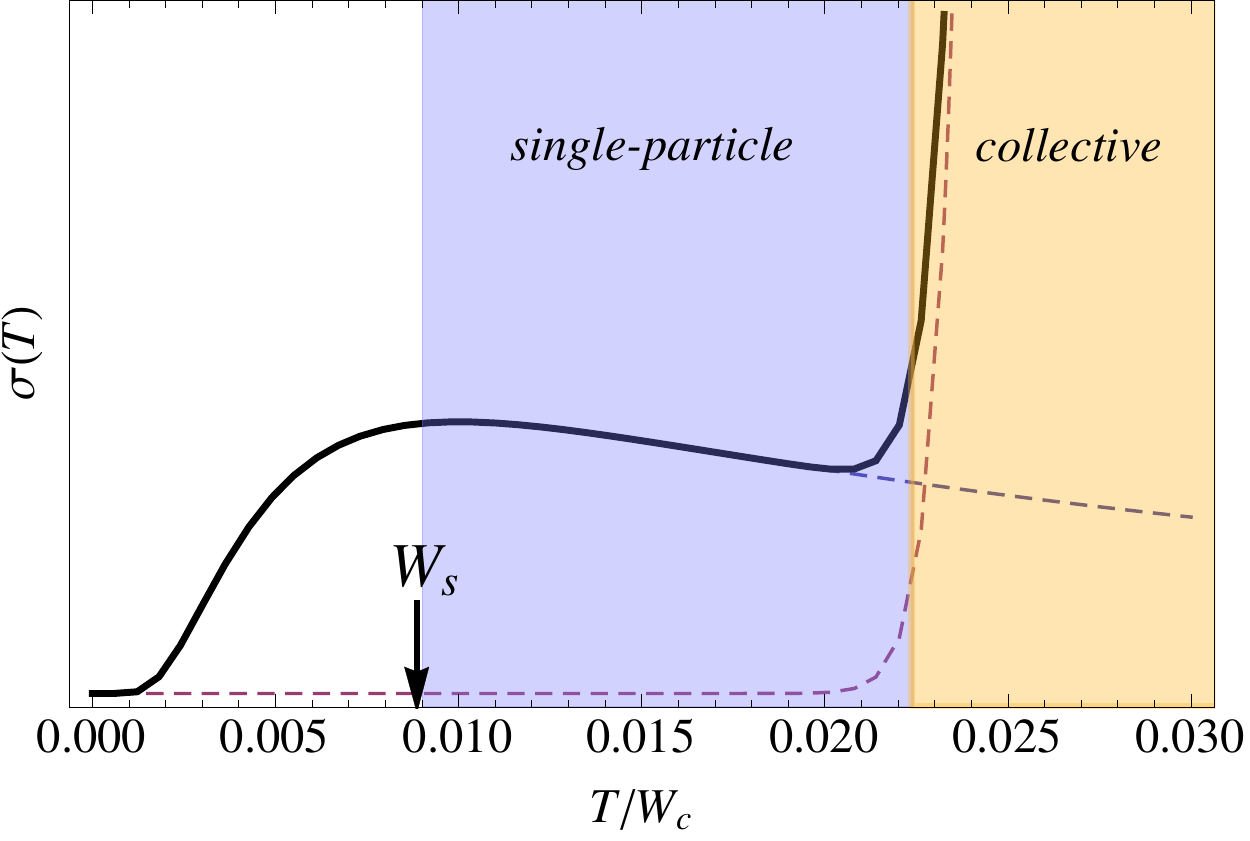}
\caption{Low-temperature d.c. charge conductivity $\sigma(T)$ of strongly interacting spinful chains, plotted for parameters {$W_s = 0.01 W_c$,  $\xi_p = 10$, $\xi_q=5$}.  
At relatively high temperatures in the SILL regime (i.e., $W_s \log(W_c/W_s) \alt T \alt W_c$), the dominant contribution to $\sigma(T)$ comes from collective rearrangements, and is activated (beige region). 
At relatively low temperatures in the SILL regime (i.e., $W_s \alt T \alt W_s \log(W_c/W_s)$, single-particle hops dominate, and $\sigma(T) \sim 1/T$ (blue region). The thick line shows the total d.c. conductivity; the single-particle and collective contributions are plotted as dashed lines. 
The behavior of $\sigma(T)$ at still lower temperatures, $T < W_s$ (i.e., in the spinful Luttinger-liquid regime rather than the SILL regime), is outside the scope of this work. We expect that the conductivity here is due to conventional hopping mechanisms and drops rapidly to zero. An appreciable regime of non-monotonic behavior exists when $\log(W_c/W_s) \gg 1$, i.e., whenever there is a well-defined SILL regime.}
\label{fig:sigmadc}
\end{center}
\end{figure}

It is straightforward to extend the previous analysis from relaxation to hopping transport. Because $W_s \ll T$, any pair of states or configurations within $W_s$ in energy automatically have an energy separation much less than $T$. It is therefore unnecessary to optimize over activation barriers (as in the standard variable-range hopping analysis~\cite{ahl}). Rather, the range over which hopping takes place is determined by the spacing between sites (or configurations) that are within $W_s$ in energy; the associated rates were computed in the previous section. Accordingly the d.c. conductivity is given, up to logarithmic corrections, by

\beq\label{qhop}
\sigma_{\mathrm{d.c.}} \simeq \frac{1}{T} (\Gamma_{\mathrm{sp}} + \Gamma_{\mathrm{int}}),
\eeq
and its overall temperature-dependence is nonmonotonic: it transitions from activated behavior at $T \agt W_s \log(W_c/W_s)$ to a $1/T$ growth at lower temperatures down to $T \sim W_s$. At still lower temperatures, presumably the d.c. conductivity vanishes again, but this is outside the regime of validity of our analysis (this regime is explored, e.g., in Ref.~\cite{yashenkin}). The various regimes are plotted in Fig.~\ref{fig:sigmadc}.

\section{Energy transport}\label{sec:energy}

Three separate channels exist for energy transport: the charge carriers (instantons) discussed in the previous section; spins; and neutral phonon-like excitations. The energy carried by spin and charge carriers is straightforward to estimate, but the contribution due to phonons is more nontrivial. {In this section, we discuss the first two of these, then estimate the phonon contribution. Comparing the three then permits us to establish regimes in which each is dominant.}

\subsection{Spin- and charge-based contributions}

Spin excitations diffuse with a diffusion constant $\sim W_s$, and each such excitation carries $\sim W_s$ of energy. It follows directly that the energy conductivity via spin excitations goes as 

\beq\label{ks}
\kappa_s(T) \sim W_s^3 / T^2.
\eeq
The charge-transport contribution to the energy conductivity is related to the charge conductivity~\eqref{qhop} by a Wiedemann-Franz law

\beq\label{kinst}
\kappa_{\mathrm{inst}} \simeq \sigma_{\mathrm{d.c.}} T; 
\eeq
thus it is activated at high temperatures and \emph{constant} at low temperatures. (More precisely, this contribution has a plateau for temperatures such that $W_s \alt T \alt W_s \log(W_c/W_s)$. At still lower temperatures, our SILL-based description does not apply, and on general grounds we expect the thermal conductivity to decrease rapidly to zero (Fig.~\ref{fig:sigmadc}).)

\subsection{``Foliated'' Variable-range hopping for phonons}

\begin{figure}
\includegraphics[width=\columnwidth]{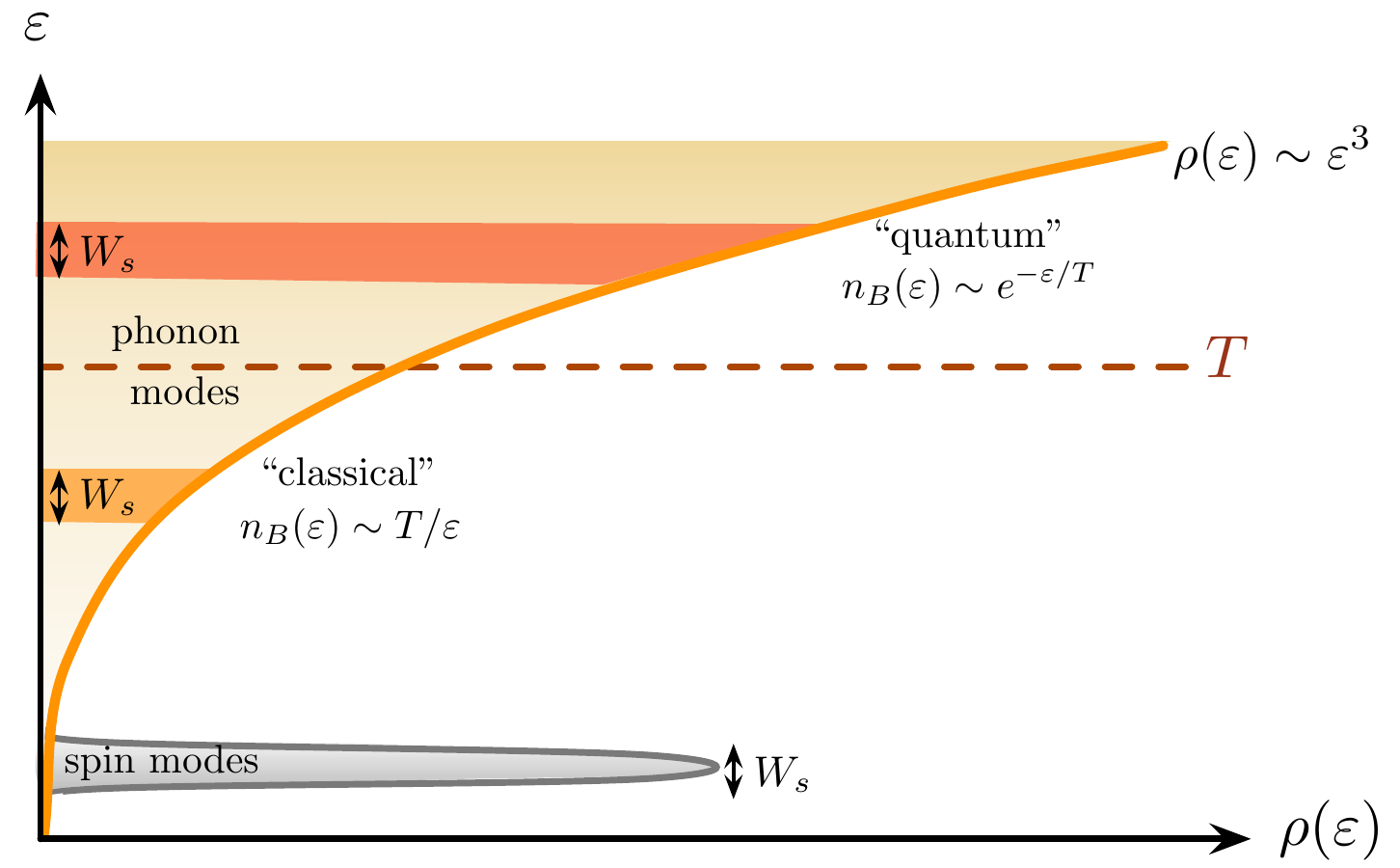}
\caption{\label{fig:foliation}``Foliated'' phonon density of states. Owing to the narrow bandwidth $W_s$ of the spin modes, spin-flip assisted boson hopping can only occur within a narrow `shell' of width $W_s$. This `foliation' leads to an emergent approximate conservation law for phonons within a particular shell. Phonons in shells centered at energy $\varepsilon \ll T$ ($\varepsilon \gg T$) are effectively classical (quantum), with occupancy $n_B(\varepsilon)\sim T/\varepsilon$ ($n_B(\varepsilon)\sim e^{-\varepsilon/T}$.)}
\end{figure}

Phonons are not conserved, so in most contexts it does not make sense to talk about their \emph{hopping} conductivity. A peculiarity of the present system, which makes phonon hopping a physically relevant channel, is the $W_s \ll T$ limit. In this limit, phonons with energy $\agt W_s$ cannot be created or destroyed at low order in the spin-charge coupling. Such phonons are extremely rare due to the vanishing phonon density of states at zero energy. Instead, the dominant phonons (which have energy $\sim T$) hop among modes that are separated by $\alt W_s$. Thus the space of phonons is ``foliated,'' (Fig.~\ref{fig:foliation}) with phonons predominantly hopping within a narrow energy range. (Interactions change this picture somewhat, as we shall discuss below.) To a good approximation (i.e. up to an energy resolution $\sim W_s$) we can consider each ``foliation'' separately.  The effective thermal conductance between two localized bosonic states $i, j$  located $R_i, R_j$ and belonging to the same foliation (i.e., for $\varepsilon_i\sim \varepsilon_j \sim \varepsilon$ to precision $W_s$) is given by (see Appendix~\ref{app:phononVRH} for details),
\be
K_{ij}(\varepsilon) \sim \frac{g^2}{W_s} \frac{\varepsilon^2}{T^2} e^{-\frac{2|R_i -R_j|}{\xi_p}} n_B(\varepsilon)[1+n_B(\varepsilon)]\label{eq:Kijfoliated}
\ee
where we have taken the spin-flip density of states  (assumed constant) to be $\nu^0_s \sim 1/W_s$, $g$ is again the spin-charge coupling, 
 $\xi_{\text{eff}}(\varepsilon)$ is the effective localization length at energy $\varepsilon$, and we drop prefactors of order one.
As noted above, the foliation of the energy spectrum leads to an approximate conservation law: there is little energy transfer between the different bands, so that we may simply consider a set of distinct hopping problems, and argue that the one with the largest thermal conductance dominates the rest. Within each energy window, the problem thus reduces to determining the effective thermal conductance of the random thermal resistor network with resistances $K_{ij}^{-1}$.  The broad distribution of the $K_{ij}$s (even at a fixed energy $\varepsilon$) permits us to argue that the {scaling of the effective phonon thermal conductivity $\kappa_{\text{ph}}$} is given by the critical $K_c$ at the percolation threshold; bonds with $K_{ij} > K_c$ fail to percolate and cannot contribute to the conductance across the whole sample, whereas those with $K_{ij}<K_c$ are shorted out by the percolating backbone. This procedure can be implemented numerically quite straightforwardly; however, we eschew this in favor of an analytical estimate that is sufficient to obtain the scaling of $K$ with temperature.

Before proceeding, we must estimate the typical real-space distance between bosonic modes at energy $\varepsilon$. The density of states of these modes may be approximated as $\rho(\varepsilon) \approx \frac{1}{c W_c} \left(\frac\varepsilon{W_c}\right)^3$ where $c$ is an $O(1)$ constant; from this, we see that the typical spacing between levels in the energy window $(\varepsilon, \varepsilon + W_s)$ is given by {$R_{\text{eff}}(\varepsilon) \sim 
c  \frac{W_c}{W_s}\left(\frac{W_c}\varepsilon\right)^3$}. Thus, we may rewrite (\ref{eq:Kijfoliated}) as 
\be
K_{ij}(\varepsilon) \sim\frac{g^2}{W_s} \frac{\varepsilon^2}{T^2} e^{-c \left[{W_c}^4/(\varepsilon^3 W_s)\right]} n_B(\varepsilon)[1+n_B(\varepsilon)]\nonumber\\\label{eq:KijfoliatedDOS}
\ee
where we have absorbed all numerical factors in the exponent by redefining the constant $c$.  The Bose factors that enter the expression for $K_{ij}(\varepsilon)$ simplify in two limits: the ``classical'' case when $\varepsilon \ll T$, and the ``quantum'' case when $\varepsilon \gg T$. We now discuss each in turn.

\subsubsection{Classical Regime}
In the classical regime, we have $n_B(\varepsilon) \approx T/\varepsilon \gg 1$, so that the typical thermal conductance of a foliation around $\varepsilon$ is
\be
K^{\text{cl}}_{ij}(\varepsilon)\sim\frac{g^2}{W_s}  e^{-c \left[{W_c}^4/(\varepsilon^3 W_s \xi_p)\right]} \label{eq:Kijfoliatedclass}
\ee
and we assume this form is valid up to $\varepsilon \sim T$, where the classical-quantum crossover occurs. Clearly, the states with $\varepsilon \ll T$ will have extremely suppressed conductances, so that the dominant classical channel is obtained right at the crossover scale. Note that the classical processes are not really `variable range': there is no tradeoff between distance and energy, and hopping always occurs to the nearest neighbor site within the same foliation. The corresponding thermal {conductivity} is
\be
\kappa^{\text{cl}}_{\text{ph}}   \sim\frac{g^2}{W_s}  e^{-c \left[{W_c}^4/(\varepsilon^3 W_s \xi_p)\right]} \label{eq:Kc_class}.
\ee
This is subleading relative to the contribution from the quantum channels (see below).

\subsubsection{Quantum Regime}
In the quantum regime, we have $n_B(\varepsilon) \approx e^{-\varepsilon/T}\ll 1$, leading to a typical conductance
\be
K^{\text{q}}_{ij}(\varepsilon) \sim\frac{g^2}{W_s}  \frac{\varepsilon^2}{T^2} e^{-c \left[{W_c}^4/(\varepsilon^3 W_s \xi_p)\right]  -\varepsilon /T}.\label{eq:Kijfoliatedquant}
\ee
We optimize the exponent among the classical channels with $\varepsilon \gg T$, and find that the dominant channel is at the energy 
\beq\label{epsilonc}
\varepsilon_c = W_c \left(\frac{3 c T}{W_s \xi_p} \right)^{1/4}.
\eeq 
The expression~\eqref{epsilonc} is only meaningful if $\varepsilon_c \alt W_c$, which is the case when $c T \alt W_s \xi_p$ (i.e., for relatively low temperatures in systems with relatively weak disorder). In this regime, the thermal conductivity from the dominant channel is
\be\label{phononfinal}
{\kappa}^{\text{q}}_{\text{ph}} (T) \sim\frac{g^2}{W_s}  a\frac{W_c^2}{\sqrt{W_s T^3}} e^{ - b W_c/(\xi_p W_s T^3)^{1/4}}
\ee
with $a = (3c/4)^{1/2}$ and $b = 7/3 (3c/4)^{1/4}$. Note that this dominates the classical contribution~\eqref{eq:Kc_class}. In the opposite limit of small $\xi_p$ or high $T$, the dominant channels are those at the highest available energies $\sim W_c$. 
The temperature dependence in this limit is \emph{activated}, although the precise rate is outside the scope of the present work (as the relevant modes are not in the SILL regime). 
Therefore we conclude that the thermal conductance due to phonons is given by Eq.~\eqref{phononfinal}, provided that the temperature is low and $\xi_p$ is large.

So far in this analysis, we have assumed that single-phonon hops dominate over multi-phonon rearrangements. This assumption holds because the dominant phonon channels (as computed above) have energies that are much higher than $T$. Therefore, such excitations are sufficiently dilute that interaction effects are expected to be subleading.

\subsection{Evolution of $\kappa$ with temperature}

\begin{figure}[tb]
\begin{center}
\includegraphics[width = 0.45\textwidth]{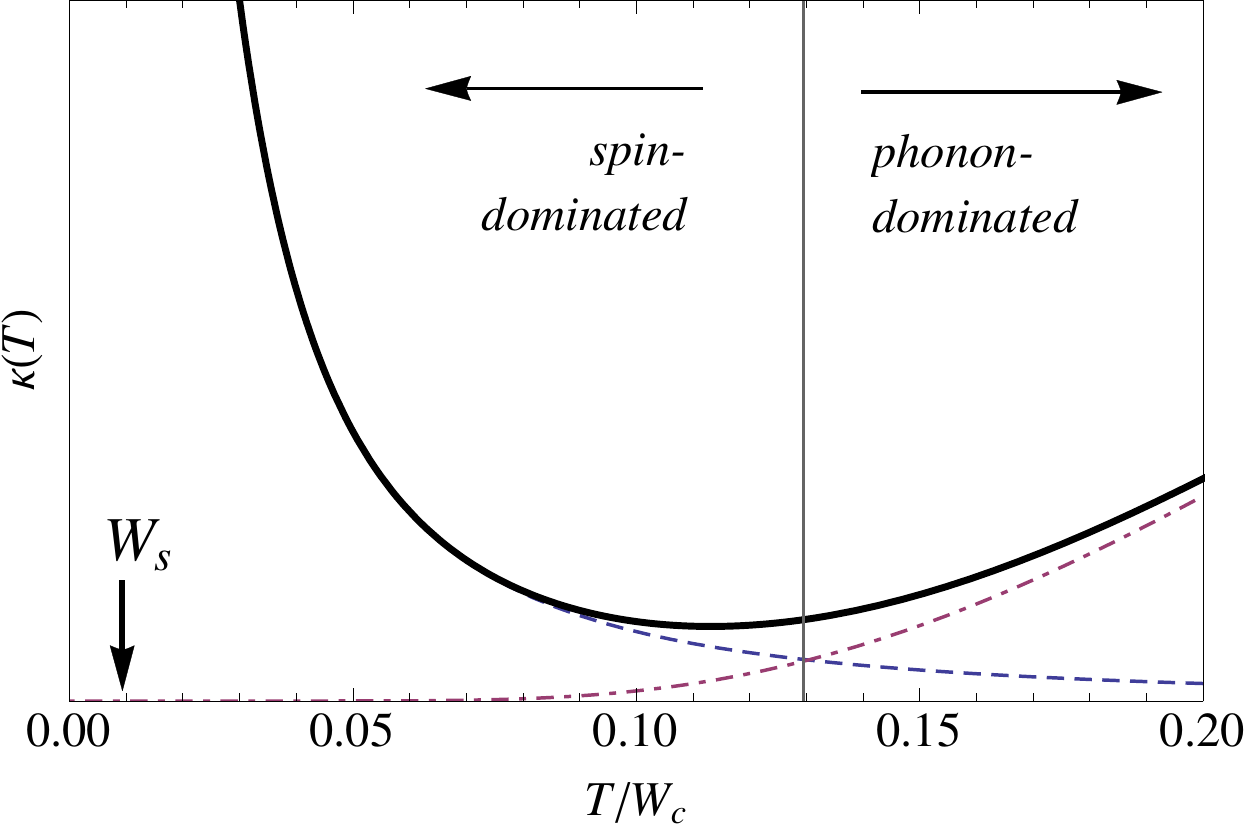}
\caption{Thermal conductivity in the SILL regime, plotted for parameters $W_s = 0.01 W_c$, $g = 0.1 W_s$, $\xi_p = 10$. For these parameters, the instanton contribution is always subleading; instead, there is a crossover from phonon-mediated energy transport (dash-dotted line) at relatively high temperatures to spin-mediated energy transport (dashed line) at relatively low temperatures. The thick line shows the behavior of the total thermal conductivity, that peaks at $T\sim W_s$. At temperatures below $W_s$, the thermal conductivity should decrease and $\kappa(T\rightarrow0) =0$; we do not discuss the details of this behavior here as it lies outside the SILL regime. For stronger disorder or weaker spin-charge coupling, the crossover temperature $T^*$ (gray vertical line) increases.}
\label{kdc}
\end{center}
\end{figure}

The three contributions to thermal conductivity at temperature $T$ are listed in Eqs.~\eqref{ks}, \eqref{kinst}, and \eqref{phononfinal}. The overall temperature-dependence of $\kappa(T)$ implied by these is as follows. At temperatures that are not much larger than $W_s$, the thermal conductivity is dominated by spin excitations, which propagate \emph{fastest} but carry the least energy per excitation. At higher temperatures, i.e., at temperatures close to $W_c$, the other channels can in principle dominate because each excitation in these channels (though slower-moving) carries more energy $\agt T$. In general, there will be a crossover between spin and phonon channels at a temperature set by

\beq
\frac{W_s^3}{(T^*)^2} \sim\frac{g^2}{W_s}  a\frac{W_c^2}{\sqrt{W_s (T^*)^3}} e^{ - b W_c/(\xi_p W_s (T^*)^3)^{1/4}}
\eeq
This equation has \emph{no} solutions for $W_s \alt T \alt W_c$ unless $\xi_p$ is sufficiently large; however, for sufficiently large $\xi_p$ (i.e., weak disorder) there is a temperature regime in which the phonons dominate over the spins. Analogous estimates suggest that instantons never dominate energy transport, as they are always subleading either to spins or to phonons. The resulting crossover is shown in Fig.~\ref{kdc}: in general, the d.c. thermal conductivity has a \emph{minimum} at temperatures between $W_s$ and $W_c$, because at these temperatures the charge degrees of freedom are essentially frozen out whereas the spin degrees of freedom contribute weakly to response because they are at infinite temperature.

\subsection{Finite-frequency thermal conductivity}

Finally, we briefly remark on the a.c. thermal conductivity at finite temperature. We expect this to be dominated by phonons, because they are much more weakly localized than instantons (assuming $\xi_q \ll \xi_p$). Let us again assume that $W_s \ll \omega \ll T$; thus the spin sector does not respond and can be neglected. The ``foliated'' analysis of the previous sections can be reprised but with $\omega$ playing the role of $W_s$. Thus $\omega$ determines a length-scale $x_\omega = \xi_p \log(W_c/\omega)$. Consider a particular foliation centered at energy $\varepsilon$. The spacing between states in this foliation is $r_\varepsilon \sim W_c^4/(W_s \varepsilon^3)$, and the fraction of occupied states is $\exp(-\varepsilon/T)$. Thus the a.c. thermal conductivity from a particular foliation is

\beq
\kappa_\varepsilon(\omega) = T (\omega r_\omega)^2 \exp(-\varepsilon/T) r_\omega/(W_c r_\varepsilon).
\eeq
This is maximized for $\varepsilon \sim T$, and so the a.c. thermal conductivity goes (up to logs) as

\beq
\kappa(\omega, T) \sim \omega^2 T^4.
\eeq

\section{Discussion}\label{sec:disc}

We have argued that the concept of ``super-universality'' breaks down for the dynamics of isolated spin-incoherent Luttinger liquids in the presence of disorder. Instead, the spin exchange timescale governs charge and energy dynamics. We have estimated the charge and energy conductivities and shown that they exhibit multiple regimes: charge transport at relatively high temperatures is due to many-body resonances, whereas at low temperatures (but still in the SILL regime) it is due to single-particle hops. Energy transport, meanwhile, is due to spin excitations at low temperatures and (for sufficiently weak disorder) due to phonons {(collective charge modes)} at higher temperatures. {Both transport coefficients evolve \emph{non-monotonically} with temperature in the SILL regime (Figs.~\ref{fig:sigmadc}, \ref{kdc}).} 

Our results generalize readily to interacting two-component systems under the following conditions: 

(i)~one of the components has a much smaller intrinsic energy scale (i.e., bandwidth) than the other, but is also subject to much weaker disorder; 

(ii)~the coupling between the two components is weak compared with the intrinsic energy scale of either. 

In the case of the SILL, which we have focused on so far, condition (i) is guaranteed by strong interactions whereas condition (ii) is a consequence of spin-charge separation. However, similar two-component systems can also be implemented, {\it e.g.}, using two-leg ladders~\cite{hyatt, yao2014quasi}, working with two or more species of particles with a large mass ratio~\cite{groverfisher},  or using weak transverse hopping in lieu of the ``spin''~\cite{bordia}. {Existing finite-size numerical studies of such systems~\cite{hyatt, mondaini} are qualitatively consistent with our conclusions; however, these  are reliable in the \emph{strongly disordered} limit, whereas our calculations are most controlled in the complementary limit of \emph{weak} disorder. }

Our results apply directly to a number of {spinful} solid-state  systems, with predominantly short-range exchange coupling, as well as to ultracold atomic gases. 
However, for experiments with semiconductor nanowires~\cite{auslaender2005, steinberg2006}, some of our results will be modified because of the power-law tail of the Coulomb interaction. In particular, rather than being exponentially localized, phonons will only be power-law localized (with tails falling off as $1/x^3$~\cite{yao2013}). One expects the d.c. conductivity in this regime to go as a power-law of the temperature, with the exponents depending on the observable as well as the power-law of the interaction: for instance, for Coulomb interacting electrons in 1D the d.c. charge conductivity $\sigma_{\mathrm{d.c.}}  \sim T^2 W_s^3 / W_c^5$. 
Since our predictions involve transport, they can be tested by standard conductivity measurements in solid-state settings~\cite{auslaender2005, steinberg2006}. Our predictions are straightforward to test in transport experiments or quench dynamics involving ultracold atoms~\cite{demarco, esslinger, mbmott}: the predictions for energy transport can be explored in cold atomic systems, {\it e.g.}, using the local thermometry scheme in Ref.~\cite{bloch:higgs}. Using this method, the a.c. thermal conductivity can also be extracted from the time-dependent autocorrelation function of the energy density. 

We now briefly discuss how our results are modified when conditions (i) and (ii) above fail. First, we consider the failure of condition (ii): 
for instance, in the two-leg ladders of Refs.~\cite{hyatt, yao2014quasi}, or in the SILL at relatively strong disorder, where spin and charge are not cleanly separated. In this case, a crucial distinction exists between systems in which the full Hamiltonian obeys $SU(2)$ symmetry and those in which it does not, {\it e.g.}, generic two-component systems or spin-orbit coupled systems. In the absence of $SU(2)$ symmetry, the charge sector can \emph{localize} the spin sector~\cite{MBLprox}, so that the full system exhibits a form of asymptotic many-body localization~\cite{2013arXiv1309.1082S, yao2014quasi} (although it is unclear at present whether such asymptotic localization is stable against rare-region effects~\cite{deroeck2014, mobilityedge16}). 
On the other hand, in the presence of $SU(2)$ symmetry, it appears~\cite{gurariechalker, QCGPRL} that the spin sector is protected against many-body localization. Thus, we expect our analysis to extend to the case of intermediate or strong spin-charge coupling for $SU(2)$ symmetric systems, {at least qualitatively}. {However, our treatment of the spin sector as being thermal but otherwise featureless might fail here. For instance, equilibrium spatial fluctuations in the charge density will lead to large spatial fluctuations in $W_s$, and regions of anomalously small $W_s$ might act as bottlenecks for hopping transport as discussed in Ref.~\cite{sggriffiths2}.}

\begin{figure}[t]
\begin{center}
\includegraphics[width = 0.48\textwidth]{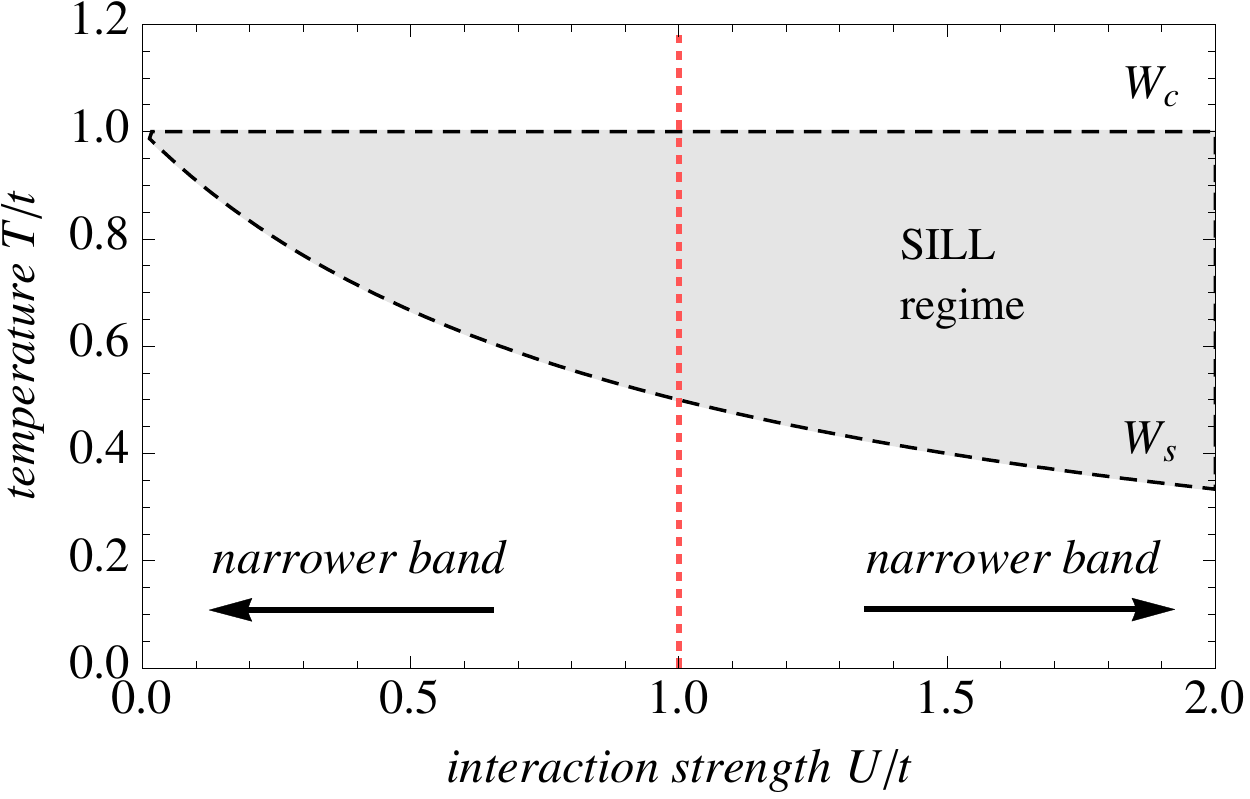}
\caption{Evolution of the SILL regime (shaded), and the characteristic bandwidth of the spin bath, for the disordered Hubbard model in one dimension. As interactions increase, the spin and charge bandwidths $W_c, W_s$ separate and an intermediate temperature regime (i.e., the SILL regime) opens up. The effective bandwidth of the spin bath, which mediates charge relaxation, is governed by $U$ at weak coupling and by $t^2/U$ at strong coupling; thus it becomes narrow (and the relaxation timescales diverge) in both limits.}
\label{zoomout}
\end{center}
\end{figure}

Finally we comment on the crossover between the strongly interacting systems considered here and the weakly interacting limit of Ref.~\cite{BAA, Gornyi} (note that Ref.~\cite{BAA}, like most of the extant literature, considered spinless fermions). For concreteness we consider the Hubbard model, and ignore spin-orbit coupling. In the noninteracting disordered problem, all orbitals are localized; the $N$-particle ground state (for even particle number) involves doubly occupying the lowest $N/2$ orbitals, and has no spin degeneracy. However, a state at finite energy density generically has a number of singly occupied orbitals, each of which is spin degenerate. For weak interactions, exchange interactions lift these spin degeneracies. If one imagines ``freezing'' the charges in a particular configuration of orbitals, the resulting spin Hamiltonian will have random exchange couplings (inherited from the positional randomness of the orbitals) but will, crucially, respect $SU(2)$ symmetry. This symmetry prevents localization in the spin sector. Thus, spins will thermalize even in this putative fixed charge background, and will thermalize the charges as well. In this low-temperature limit, the density of singly occupied orbitals goes as $T/E_F \sim T/W_c$; consequently the exchange coupling between them (which sets the bandwidth of the spin bath) will go as $U \exp(-(E_F/T)/\xi)$ where $\xi$ is the single particle localization length. Thus the physics of relaxation through a narrow-bandwidth spin bath also applies in the weak-coupling regime. However, the charge and spin temperatures are essentially the same at weak coupling (as $W_c \simeq W_s$), so the unusual non-monotonic transport signatures discussed in this work will not be present there. Fig.~\ref{zoomout} summarizes the various regimes.  

In closing, we observe that, for reasons described in Section~\ref{sec:scales}, the disordered, {isolated} SILL is {\it not} a many-body localized system. {It has two intrinsic channels for thermalization, viz. the spin bath that we have focused on, as well as the high-energy charge modes, which are thermal in the strongly interacting, weakly disordered limit where our calculations are controlled}. Thus the SILL is in fact an ergodic system --- for instance, we expect that its eigenstates are volume-law entangled, and that observables computed in single eigenstates at finite energy density will exhibit thermal behavior. Nevertheless, we have shown that the transport and dynamics show many features that are most easily understood by beginning with an MBL system and {adding perturbations that thermalize it}. In this sense, the SILL provides a new example of a {thermal} system whose dynamics are fruitfully addressed from the MBL perspective. We anticipate that there are other situations where such phenomenology emerges, and that similar ``MBL-controlled'' {crossovers} may be surprisingly ubiquitous, particularly in low-dimensional disordered systems.


 \vspace{10pt}\noindent \textbf{\textit{Acknowledgements.}} We thank B.~Bauer, A.L.~Chernyshev, T.~Grover, D.A.~Huse, F.~Huveneers, M.~Knap, I.~Lerner, R.~Nandkishore, V.~Oganesyan, A.C.~Potter, and S.L.~Sondhi for helpful discussions.  This research was  supported in part by the National Science Foundation under Grants No. NSF PHY11-25915 at the KITP (SAP, SG) and CAREER Award DMR-1455366 (SAP). S.G. acknowledges support from the Burke Institute at Caltech. {We acknowledge the hospitality of the KITP and highways US-101, CA-134, CA-118, and CA-126, where portions of this work were carried out.}
  \vspace{10pt}
\begin{appendix}

\section{Phonon variable-range hopping}\label{app:phononVRH}
In this appendix, we provide details of of the Gaussian-sector VRH calculation. As in the main text we consider a set of localized bosonic modes at random positions $i$, with  random energies $\varepsilon_i>0$ (the positivity of constraint is because the Gaussian sector describes excitations above the pinned CDW ground state), distributed according to the density of states $\rho(\varepsilon)$.

\begin{figure}[h]
\includegraphics[width=\columnwidth]{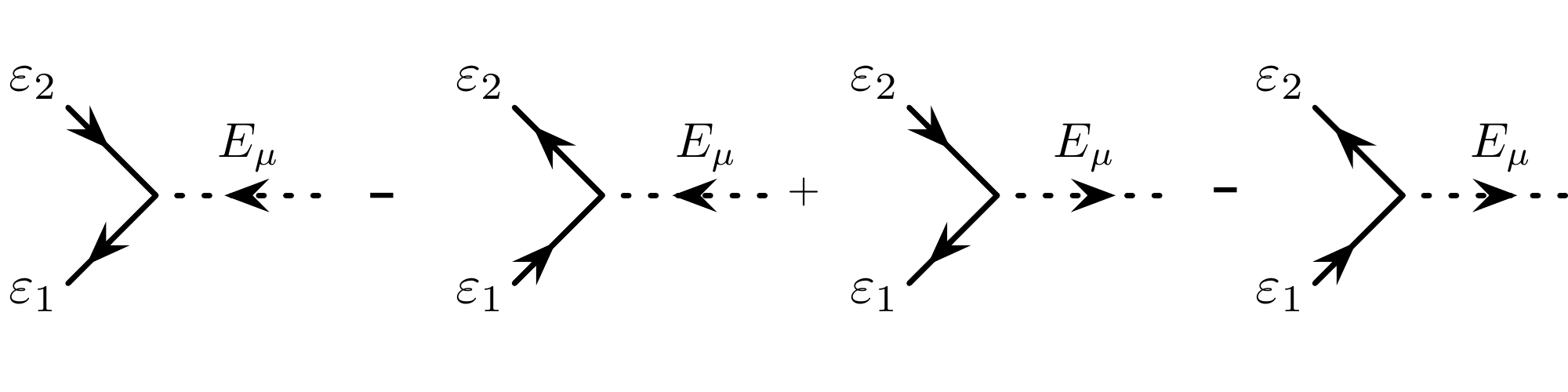}\caption{\label{fig:bose_diags} Diagrams for one-spin-flip absorption and emission  processes that contribute to the transition rate for localized bosonic states.}
\end{figure}

The change in the distribution function of level $1$ due to transitions to and from level $2$ is obtained from Fermi's golden rule as
\begin{widetext}\be
\partial_t n(\varepsilon_1) &=& 2\pi g^2\, e^{-2|R_i -R_j|/\xi_p}\sum_{\mu} \left[\delta(\varepsilon_1-\varepsilon_2-E_\mu)\left\{(1+n_B(\varepsilon_1))n_B(\varepsilon_2)n_B(E_\mu)- n_B(\varepsilon_1)(1+n_B(\varepsilon_2))(1+n_B(E_\mu)) \right\}\right.\nonumber\\ & & +\left. \delta(\varepsilon_1+\varepsilon_2-E_\mu)\left\{(1+n_B(\varepsilon_1))n_B(\varepsilon_2)n_B(-E_\mu)- n_B(\varepsilon_1)(1+n_B(\varepsilon_2))(1+n_B(-E_\mu)) \right\}\right] 
\ee
\end{widetext}
where $\mu$ indexes all the single-spin-flip processes in the first expression we have approximated a common matrix element $|\mathcal{H}_{\text{el-s}}|^2 \approx g^2\, e^{-2|R_i -R_j|/\xi_p}$ 
for all single-spin-flip processes involving bosons localized on sites $i, j$, where $\xi_p$ is the pinning length.

We may perform the sum over $\mu$ by converting into an integral $\sum_\mu (\ldots) \rightarrow \int_0^\infty d\varepsilon \nu_s(\varepsilon) (\ldots)$ where $\nu_s(\varepsilon)$ is the density of states of the spin-flip processes. From this, we obtain a general formula for the transition rate from state $i$ to state $j$,
\begin{widetext}\be
\Gamma_{i j}^0(\varepsilon_i, \varepsilon_j, R_i, R_j) &=& 2\pi g^2\, e^{-2|R_i -R_j|/\xi_p}
 \nu_s(|\varepsilon_i-\varepsilon_j|) n_B(\varepsilon_i)(1+n_B(\varepsilon_j))\left\{ \begin{array}{cc} \tilde{n}_B(\varepsilon_j -\varepsilon_i), &\varepsilon_i < \varepsilon_j \\  1+ \tilde{n}_B(\varepsilon_i -\varepsilon_j), & \varepsilon_i > \varepsilon_j.  \end{array}  \right.
\nonumber\\
&\equiv& n_B(\varepsilon_i)(1+n_B(\varepsilon_j)) \gamma_{ij}^0 (\varepsilon_i, \varepsilon_j, R_i, R_j).
\ee\end{widetext}
where we have separated out the occupancy factors from the `intrinsic' transition rate $\gamma^0_{ij}$. {The occupation factors $\tilde{n}_B$ for the spin-sector excitations are taken to be those of strongly anharmonic bosonic modes, i.e., the thermal occupation of each spin mode is truncated at a number on the order of unity}.   It is readily verified that these rates satisfy the detailed balance condition $\Gamma^0_{ij} = \Gamma^0_{ji}$, required to define equilibrium in the absence of temperature and field gradients. As a next step, we need to relate the thermal conductivity to the equilibrium hopping rate $\Gamma^0_{ij}$. To that end, we imagine imposing a temperature gradient $\nabla T$, so that the sites $i$ and $j$ are at different temperatures.   While this shifts the occupancy factors by adjusting the local temperature at sites $i, j$, there is no change in the intrinsic transition rate $\gamma_{ij}^0$. This is a consequence of the Bose factors entering $\gamma_{ij}^0$ reflect the occupancy of the spin-flip mode absorbed or emitted to make up the energy difference between $\varepsilon_i, \varepsilon_j$; since the characteristic energy scale $|\varepsilon_i-\varepsilon_j|$ is set by the maximal spin-flip energy $\sim W_s$ and we have $T\gg W_s$, small variations in the temperature over distance $\sim \xi_{\text{eff}}$ may be ignored, so that we may simply compute $\gamma^0_{ij}$ at the average temperature of sites $i, j$, namely $T$.  Under these assumptions, the differential rate for boson hopping between sites $i$ and $j$ is obtained as (putting explicit temperature dependence in the occupancy factors to reflect the thermal gradient)
\begin{widetext}
\be
\Delta \Gamma_{ij} (\nabla T) &\equiv& \left.\Gamma_{ij}(\varepsilon_i, \varepsilon_j, R_i, R_j)\right|_{\nabla T} - \left.\Gamma_{ji}(\varepsilon_i, \varepsilon_j, R_i, R_j)\right|_{\nabla T} \nonumber\\
&=&  \gamma_{ij}^0 (\varepsilon_i, \varepsilon_j, R_i, R_j) n_B(\varepsilon_i, T_i)(1+ n_B(\varepsilon_j, T_j)) -   \gamma_{ji}^0 (\varepsilon_i, \varepsilon_j, R_i, R_j) n_B(\varepsilon_j, T_j)(1+ n_B(\varepsilon_i, T_i)) \nonumber\\
&=& \Gamma^0_{ij} \left[\frac{\delta n_i}{n_i^0(1+n_i^0)} - \frac{\delta n_j}{n_j^0(1+n_j^0)} \right]
\ee
\end{widetext}
where we have defined $n_i^0 \equiv n_B(\varepsilon_i, T)$, and  $\delta n_i  \equiv n_B(\varepsilon_i, T_i) - n_B^0$. Assuming the linear-response regime, we may take $T_{i,j} \approx T \pm \frac{\vec{R}_{ij}}{2}\cdot \vec{\nabla} T \equiv T \pm \frac{\delta T}2$, where $\vec{R}_{ij} \equiv R_i - R_j$. With this parametrization, we have, after a little work,
\be
\Delta \Gamma_{ij} (\nabla T) = \frac{\varepsilon_i +\varepsilon_j}{2 T^2} \Gamma_{ij}^0\times \left(  \vec{R}_{ij}\cdot \vec{\nabla} T \right).
\ee
In order to obtain the energy current, we must multiply this number current by the typical energy transported in the tunneling process, which we take to be the average energy of sites $i,j$, yielding
\be
I^{(Q)}_{ij} = \frac{(\varepsilon_i +\varepsilon_j)^2}{4 T^2} \Gamma_{ij}^0\times \left(  \vec{R}_{ij}\cdot \vec{\nabla} T \right).
 \label{eq:Jq}
\ee
Note that the expression in parentheses is the net temperature difference between the two sites; thus, the remainder of the RHS of (\ref{eq:Jq}) is the thermal conductance between sites $i, j$,
\be
K_{ij}=  \frac{(\varepsilon_i +\varepsilon_j)^2}{4 T^2} \Gamma_{ij}^0(\varepsilon_i, \varepsilon_j, R_i, R_j).
\ee
Two observations allow us to simplify the expression above. First, in the usual electron variable-range-hopping computation, the density of hopping levels $\rho(\varepsilon)$ is treated as roughly constant, in contrast to the scaling $\rho(\varepsilon) \sim \varepsilon ^\gamma$ appropriate to the quantized Gaussian fluctuations of the CDW. Second, the phonon bath invoked in those treatments is assumed to be able to absorb and emit at any frequency: the ``perfect bath" limit. In essence, this allows us to take $|\varepsilon_i -\varepsilon_j| \gg T$ when computing the intrinsic rate. Here, in contrast, we have a narrow bath, and therefore the spin-flip density of states vanishes for large energy differences $ |\varepsilon_i -\varepsilon_j| \gtrsim W_s$, and we are working in the regime where $W_s \ll T$. As a consequence, $\Gamma_{ij}^0$ vanishes unless  $|\varepsilon_i -\varepsilon_j| \lesssim W_s$; since $\rho(\varepsilon) \sim \varepsilon^\gamma$ over a range $W_c \gg W_s$, it follows that we must consider a sequence of VRH problems within energy bands of width $W_s$ and with a density of localized states given by $\rho_s$. In other words, this is the `foliation' discussed in the main text: since $\nu_s(\varepsilon_i-\varepsilon_j) \approx \frac{1}{W_s} \Theta(W_s - |\varepsilon_i-\varepsilon_j|)$, both levels $\varepsilon_i, \varepsilon_j$ are at approximately the same energy (within the resolution of the bath bandwidth) for all the factors on the RHS. As a consequence, within the energy resolution $W_s$, we may approximate $\Gamma^0_{ij}$ by a hopping rate %
that depends on a single energy $\varepsilon$ ,
\be
\Gamma^0_{ij} &\approx& 
2\pi g^2\nu_s^0  e^{-\frac{2|R_i -R_j|}{\xi_p}} 
n_B(\varepsilon)[1+n_B(\varepsilon)],
\ee
whence we find that the thermal conductance between $i, j$ is
\be
K_{ij}(\varepsilon) \approx 2\pi g^2\nu_s^0 \frac{\varepsilon^2}{T^2} e^{-\frac{2|R_i -R_j|}{\xi_p}}
n_B(\varepsilon)[1+n_B(\varepsilon)].\nonumber\\
\ee

\end{appendix}
\bibliography{MBL}
\end{document}